# Stress-driven dynamic evolution of core-shell structured cavities with H and He in BCC-Fe under fusion conditions

Jin Wang[a], Fengping Luo[a], Yiheng Chen[a], Denghuang Chen[a], Bowen Zhang[a], Yuxin Liu[a], Guangyu Wang[b], Yunbiao Zhao[a], Sheng Mao[b], Mohan Chen[a, b], Hong-Bo Zhou[c], Jianming Xue[a], Yugang Wang[a], and Chenxu Wang[a, *]

[a] *State Key Laboratory of Nuclear Physics and Technology, Center for Applied Physics and Technology, Peking University, Beijing, 100871, China*

[b] *School of Mechanics and Engineering Science, Peking University, Beijing 100871, China*

[c] *Department of Physics, Beihang University, Beijing, 100191, China*

* Corresponding authors. E-mail addresses: cxwang@pku.edu.cn (Chenxu Wang).

## Abstract

Understanding the dynamic behavior of microstructures formed under fusion conditions is critical for designing high-performance structural materials for fusion reactors. Under fusion conditions, cavities of core-shell structures are formed due to the interaction between irradiation-induced vacancies and H and He atoms produced via transmutation. In this study, thermodynamic analysis and molecular dynamics simulations are combined to investigate the atomic-scale mechanisms and dynamic response of core-shell cavities formed in BCC-Fe under applied stress/strain fields. The thermodynamic analysis provides both the foundational reference for cavity structures under fusion neutron irradiation and the initial configurations for atomistic simulations. Building on this framework, atomic-scale simulations demonstrate that H and He play a decisive role in the stress-strain response and the evolution of elastic-plastic deformation within the cavities. In core-shell configurations, H atoms serve a function analogous to that in He-filled cavities, synergistically interacting with He to induce cavity deformation under mechanical loading.

Fusion energy represents a frontier technology with the potential to address global energy and climate challenges, positioning it as a leading candidate for future sustainable energy. However, the realization of fusion energy is critically dependent on the development of structural materials capable of withstanding the extreme conditions within a fusion reactor [1]. Unlike fission neutrons, Deuterium-Tritium fusion neutrons not only lead to severe displacement damage, but also generate substantial amounts of hydrogen (H) and helium (He) via nuclear transmutation [2–5]. The synergistic interaction of these gases can lead to the formation of unique microstructural defects, which significantly degrade the macroscopic properties of materials [2]. In the absence of high-flux fusion neutron sources, multi-ion beam irradiation techniques [6–13] and computational modeling approaches [14–21] serve as essential tools for exploring the effects and mechanisms of displacement defects and transmutation gases under fusion conditions.

The synergistic effect of displacement damage and He generation has been found to promote cavity swelling across different temperatures [6,8,9,22], leading to degradation of mechanical properties, such as irradiation hardening and He-induced grain boundary embrittlement [23–27]. As a typical irradiation-induced microstructure, He bubbles play a critical role in determining the overall mechanical response of materials due to their dynamic characteristics [28]. Recently, significant research efforts have utilized *in situ* transmission electron microscopy (TEM) to investigate the evolution of He bubbles under both irradiation [11,29,30] and mechanical loading [28,31–33], establishing direct correlations between their behavior and macroscopic mechanical properties. At the atomic scale, molecular dynamics (MD) simulations have provided complementary mechanistic insights into these microstructural processes, thereby elucidating the physical origins of the experimentally observed phenomena. For example, some MD studies have simulated the continuous introduction of He into metallic samples, forming over-pressurized bubbles whose growth and evolution occur *via* the loop-punching mechanism, involving the emission of dislocation loops [34–38]. Similarly, other simulations have demonstrated that low-pressure He bubbles subjected to external stress can grow through dislocation-dominated yielding behavior [39–42].

Furthermore, multi-beam ion irradiation experiments have shown that H/He co-injection leads to more pronounced cavity swelling compared to He implantation [10,12], indicating the significant non-negligible role of H in irradiation-induced microstructural evolution. However, the evolution behavior of cavities induced by the interaction of irradiation-induced vacancy defects with transmutation gases (H and He) under fusion conditions remains insufficiently understood. In particular, the role of H within He-containing cavities under stress/strain fields is still unclear, partly due to the historical challenge of developing accurate and efficient interatomic potentials for the Fe–H–He system. Our recent study [19] employed a newly developed Fe–H–He interatomic potential and thermodynamic analysis to reveal the potential physical mechanisms governing the formation of cavities with core-shell structures (where He is confined at the cavity center while H accumulates at the surface) under various environments. Based on this understanding, further investigation into the dynamic response and evolution mechanisms of cavities containing both H and He atoms under stress/strain fields (irradiation-induced or externally applied loads) is fundamental for designing iron (Fe)-based structural materials with ideal irradiation resistance.

The present work first employs thermodynamic analysis to determine the most probable occupancy of H atoms in different cavity configurations. These results guide the establishment of physically representative initial structures in subsequent MD simulations. After constructing a core-shell cavity model with appropriate numbers of H and He atoms, a triaxial tensile load was applied within the MD framework to introduce a hydrostatic strain field. This loading method is commonly used to study the dynamic evolution of pre-existing cavities in alloys [39,43,44], as well as void nucleation and growth in bulk materials [45]. The interatomic interactions were described using the embedded-atom method (EAM) potential developed by Huang et al. [17], which has demonstrated both computational efficiency and accuracy in modeling the energetics of H and He atoms interacting with vacancy and self-interstitial atom clusters in BCC-Fe [17–19]. All simulations were performed using LAMMPS (Large-scale Atomic/Molecular Parallel Simulator) [46], with visualization and defect analysis conducted in OVITO [47]. Further simulation details are provided in the Supplementary Materials.

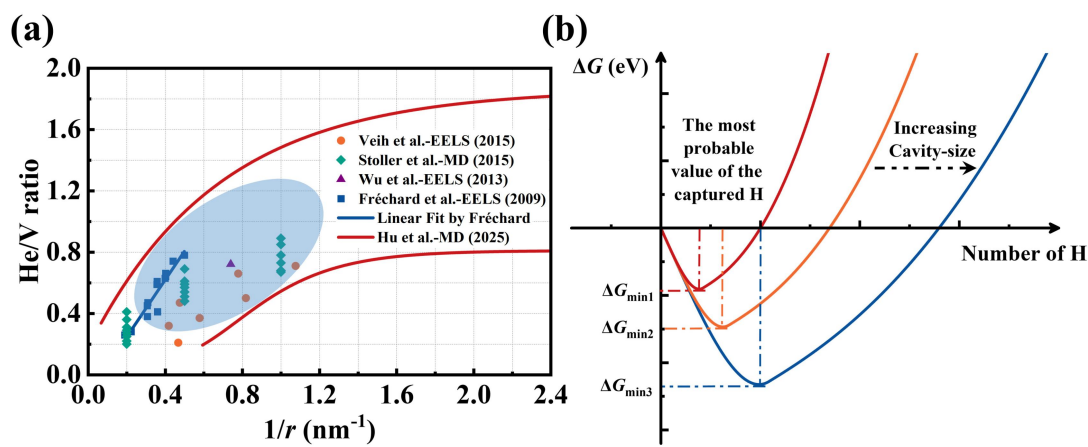

**Fig. 1.** (a) Experimental and calculated data on the He/V ratio of He bubbles formed in steels under irradiation conditions [48–52]. The cavity sizes corresponding to the experimental data symbolized by blue geometric shapes are the primary range of consideration in this study. (b) Schematic diagram for determining the most probable values of H atoms by different cavities (varying in size and He/V ratio) through calculating the change in Gibbs free energy of the system ($\Delta G$) after additional H atoms are trapped.

Fig. 1 (a) summarizes data on the He content or He/V ratio in stable He bubbles formed in steels, compiled from numerous irradiation experiments and MD simulations [48–52]. Due to discrepancies in irradiation temperature and He implantation rate, the reported He/V ratio is not a unique value but instead occupies a specific range for a given condition. MD calculations further provide boundary conditions for the He/V ratio as a function of cavity size [52]. Exceeding this critical ratio activates the loop-punching mechanism, leading to instability in both the size and structure of the cavity. Therefore, integrating the data from these sources [48–52] can establish reliable He/V ratio ranges for cavities of specific radius (*r*). The cavity sizes corresponding to the experimental data symbolized by blue geometric shapes are the primary range of consideration, as shown in Fig. 1 (a). To encompass the range of cavity structures expected under varying irradiation conditions, this study examined cavities with three representative radii and their corresponding He/V ratio ranges: 0.86 nm (He/V = 0.64 ~ 1.57), 1.43 nm (He/V = 0.28 ~ 1.26), and 2.86 nm (He/V = 0.2 ~ 0.82). For detailed analysis, specific He/V ratios were chosen from these ranges for each cavity size: 0.8, 1.2, 1.5 (*r* = 0.86nm), 0.5, 0.8, 1.2 (*r* = 1.43 nm), 0.2, 0.5, 0.8 (*r* = 2.86 nm).

After the cavity structure (He/V ratio and size) in the initial system has been well-defined, the change in Gibbs free energy ($\Delta G$) during the capture of the *i*-th H atom by the cavity could be expressed as:

$$\Delta G = \Delta H - T\Delta S \quad (1)$$

where $\Delta H$, $\Delta S$, and $T$ represent the enthalpy change, entropy change, and thermodynamic temperature during the trapping process, respectively.

Based on the previous reasonable definition [16,19,53], the enthalpy change within the system is expressed as:

$$\Delta H = -\sum_{z=2}^{m} E_b^v(0,0,z) - \sum_{y=1}^{n} E_b^{He}(0,y,m) - \sum_{x=1}^{i} E_b^H(x,n,m) \quad (2)$$

The terms $E_b^v$, $E_b^{He}$, and $E_b^H$ in the equation represent the binding energies of the vacancy, He atom, and H atom to the cavity, respectively. Here, *m*, *n*, and *i* represent the number of vacancies, He and H atoms in the formed stable cavity ($H_i He_n V_m$).

The entropy change within the system can thus be approximated as follows [19]:

$$\Delta S = -(m-1)k_B \ln \frac{C_V}{1-C_V} - nk_B \ln \frac{C_{He}}{1-C_{He}} - ik_B \ln \frac{C_H}{1-C_H} \quad (3)$$

where $C_V$, $C_{He}$, and $C_H$ represent the concentrations in the matrix of the system, respectively, and $k_B$ denotes the Boltzmann constant.

Consequently, Eq. (1) is reasonably defined as:

$$\Delta G(i,n,m) = -\left\{\sum_{z=2}^{m} E_b^v(0,0,z) + \sum_{y=1}^{n} E_b^{He}(0,y,m) + \sum_{x=1}^{i} E_b^H(x,n,m)\right\} - \{(m-1)k_B T \ln \frac{C_V}{1-C_V} + nk_B T \ln \frac{C_{He}}{1-C_{He}} + ik_B T \ln \frac{C_H}{1-C_H}\} \quad (4)$$

For a stable cavity with fixed vacancy ($C_V$) and He concentrations ($C_{He}$), the following thermodynamic criterion determines its capacity to trap H under given irradiation temperature and H concentration in the matrix ($C_H$): if the change in system free energy after capturing the ($i$+1)-th H atom is consistently lower than that after capturing the $i$-th atom, i.e., $\Delta G$ ($i$+1, $n$, $m$)−$\Delta G$ ($i$, $n$, $m$)<0, then the cavity can spontaneously trap additional H.

Therefore, the most probable number of H atoms captured by a He-containing cavity under specific irradiation conditions and cavity geometries can be determined by identifying the minimum system free energy ($\Delta G_{min}$), as shown in Fig. 1 (b). Moreover, the details of the H/V ratio formed after H atoms are trapped by the cavity under different cavity sizes and He/V ratios are presented in Table 1. Notably, the selected irradiation temperature of 723 K corresponds to the peak swelling temperature observed in steels [10,12], and the matrix H concentration of 1000 appm is representative of levels expected under high irradiation doses [2,8].

## Table 1

Thermodynamic analysis of the stable H/V ratio formed by Hydrogen (H) atoms captured within cavities of varying structures (different sizes and He/V ratios) under specific irradiation conditions.

| Cavity structures | He/V ratios | Irrad. conditions | H/V ratios |
|---|---|---|---|
| *r* = 0.86 nm | 0.80 | | 0.75 |
| | 1.20 | | 0.70 |
| | 1.50 | **Temperature 723 K** | 0.67 |
| *r* = 1.43 nm | 0.50 | | 0.48 |
| | 0.80 | | 0.46 |
| | 1.20 | **H concentration 1000 appm** | 0.43 |
| *r* = 2.86 nm | 0.20 | | 0.26 |
| | 0.50 | | 0.25 |
| | 0.80 | | 0.23 |

Based on thermodynamic results, a core-shell structured cavity [10,19] with physically realistic numbers of He and H atoms was constructed for cavities formed under specified irradiation conditions. Subsequently, triaxial tensile loading was applied to the entire system to introduce a hydrostatic strain field, as illustrated in Fig.

2 (a). The selected strain rate of $1\times10^6$ s$^{-1}$ facilitates the observation of cavity evolution over extended simulation timescales while providing atomic-scale insights relevant to high-strain-rate deformation [45]. For comparison, pure voids (He/V = 0) of corresponding sizes were also investigated.

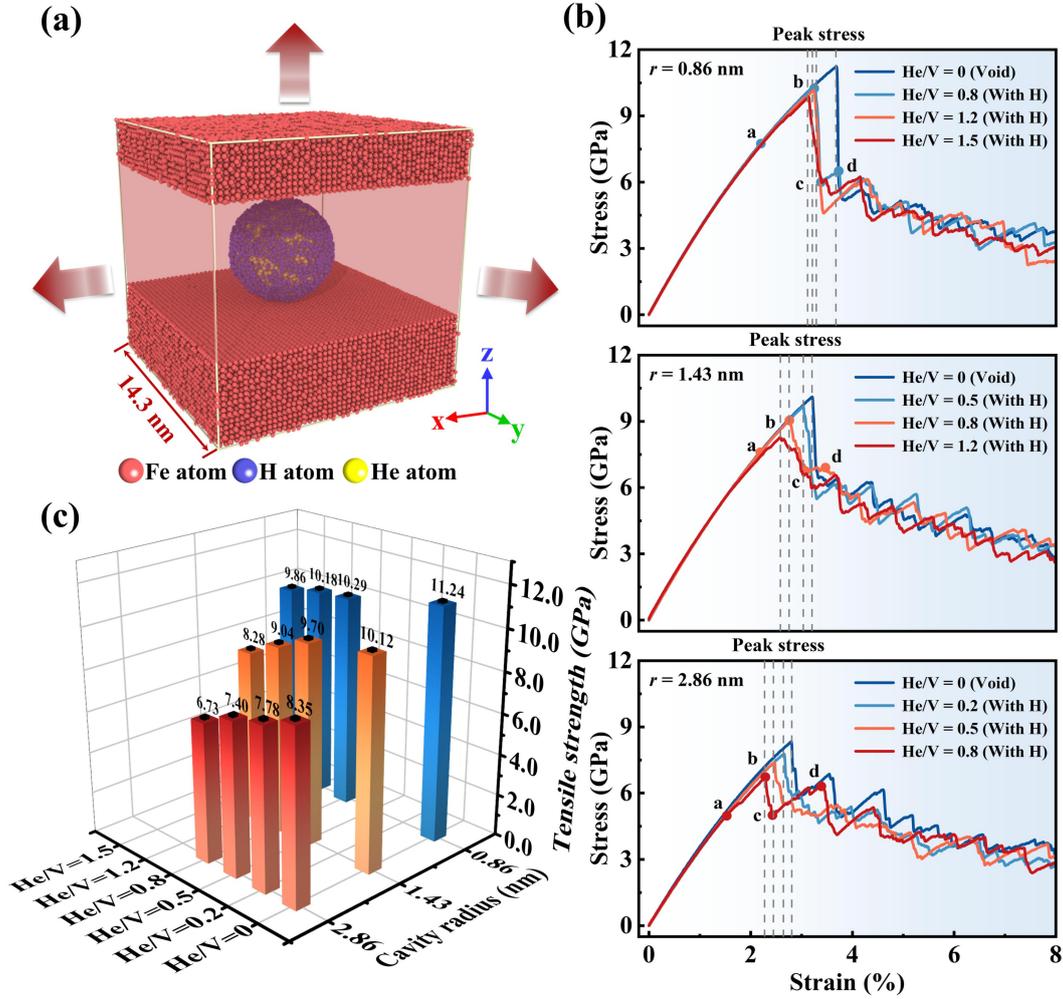

**Fig. 2.** (a) Schematic diagram of the hydrostatic strain field induced by triaxial tensile loading in BCC-Fe containing cavities with H and He. (b) Evolution of stress-strain curves in BCC-Fe with cavities of different structures at $1\times10^6$ s$^{-1}$ and 723 K. (c) Response of tensile strength (corresponding to peak stress in (b)) for these cavities.

As presented in Fig.2 (b), the stress-strain response demonstrates that H/He-filled cavities substantially modify all deformation stages relative to pure voids, a trend consistent across all cavity sizes studied. The initial elastic stage seems not to be affected by the core-shell cavity structure for all sizes, with the stress increasing linearly until reaching a peak value. However, the peak stress (tensile strength) is significantly influenced by the coexisting H and He atoms within the cavity. The tensile strength exhibits a systematic decrease with increasing He/V ratio for each cavity size, as shown quantitatively in Fig. 2 (c). Furthermore, these cavities significantly reduce the critical strain required for the bulk material to reach peak stress, and this effect is more significant as the He/V ratio within the cavities increases. After reaching peak stress, the bulk is subjected to plastic deformation, and the evolution of the stress-strain curve reveals that the plastic behavior is significantly influenced by the coexistence of H and He atoms within the cavity.

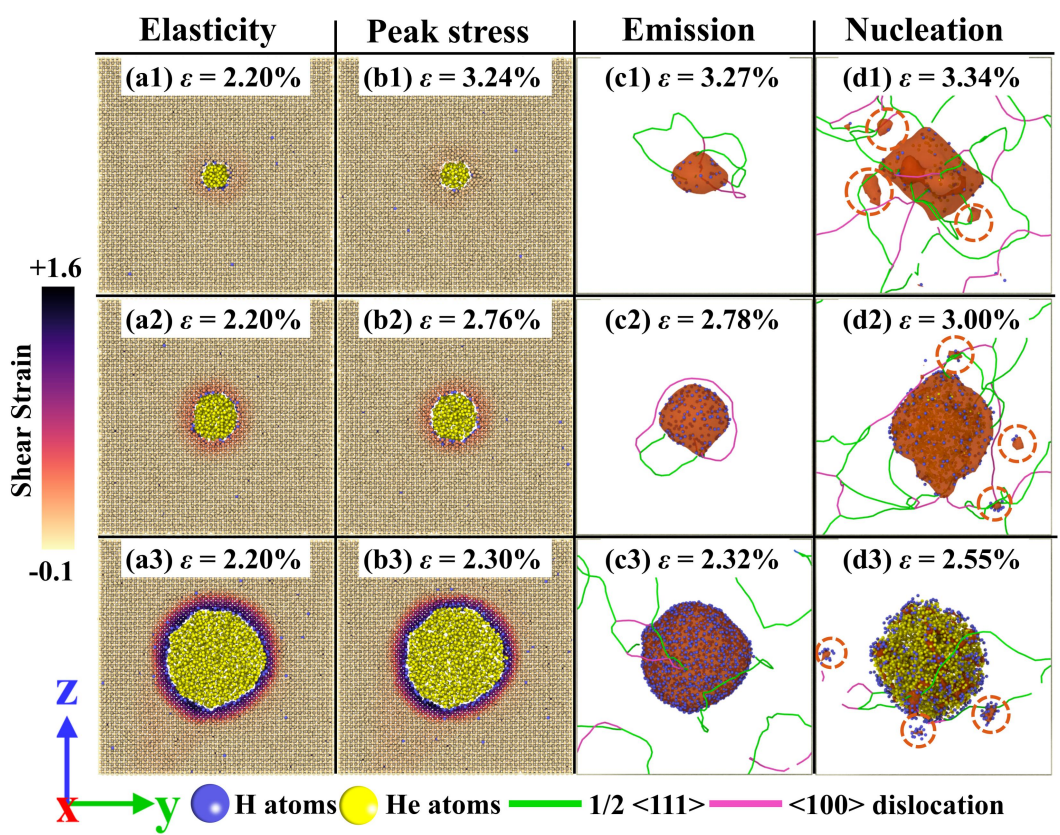

**Fig. 3.** Atomic distribution (emphasized by shear strain) and defect configuration diagrams (cavity surfaces characterized as orange by the CSM algorithm) for each size at He/V=0.8 under a hydrostatic strain ($\varepsilon$) field. The cross-sectional views of the three sizes correspond to the deformation response stages indicated by points (a), (b), (c), and (d) on the stress-strain curve shown in Fig. 2 (b). The cavities nucleated in the BCC-Fe during the plastic deformation stage are marked with orange circles.

To elucidate the underlying mechanisms, Fig. 3 provides atomic distribution and defect structure configurations at different stages of stress-strain curve evolution. These represent characteristic stages in the deformation process of cavities with different structures under hydrostatic strain fields, namely elastic deformation, peak stress, plastic deformation, and cavity nucleation. While the elastic regime of the stress-strain curve appears unaffected by the presence of H and He across all cavity sizes, the strain field surrounding the cavity is significantly intensified by the synergistic interaction of the two gases. This effect becomes more pronounced with increasing cavity sizes (as exemplified by a constant He/V=0.8), as shown in Figs 2 (b) and 3. This observation agrees with our recent finding [19] that H and He trapped in unloaded cavities likewise modify the local strain field and its interaction with other defects.

After the applied stress reached the peak stress, the sample was subjected to an elastic-plastic deformation transition. Subsequently, these cavities were observed to act as dislocation sources, and the subsequent emission of dislocations from the cavity surfaces promoted the occurrence of plastic deformation, as depicted in Figs. 3 (c1-c3). Quantitative data on dislocation density, cavity volume, and cavity number for each configuration, alongside corresponding atomic distribution and defect analysis, are provided in the Supplementary Materials. Combined with the stress-strain response shown in Fig. 2 (b), it is observed that dislocation emission is more readily activated in samples with higher He/V ratios and larger sizes. This is attributed to the internal pressure generated by He atoms at the core of the cavity, which leads to a sustained outward tensile stress on the cavity inner surface [31,40–42]. Moreover, larger cavities act as stronger stress concentration regions, thereby facilitating the induction of plastic deformation by cavities containing both H and He atoms. As dislocations continuously emit from the cavity surface, the dislocation pile-ups become potential nucleation sites for micro-voids when a certain dislocation density is reached in the bulk [54,55], further influencing the macroscopic mechanical response of the cavity-containing sample.

To clarify the role of H and its mechanism within cavities featuring a core-shell structure, further investigations employed a larger-sized cavity (r = 2.86 nm) and compared the stress-strain curve response and defect evolution between cavities containing both H and He and those containing only He atoms, as illustrated in Fig. 4. Compared to pure He bubbles, the presence of H in the cavity further influences the stress-strain evolution, as shown in Fig. 4 (a). Moreover, Fig. 4 (b) quantifies this effect by presenting the change in tensile strength (Δ$TS$) for both H/He and He-only cavities relative to pure voids of the same size. The data clearly indicate that H atoms at the cavity shell cause a marked reduction in tensile strength. For a given He/V ratio, the effect of H is evidenced not only by a reduction in the strain required for the bulk to reach its peak stress, but also by a significant decrease in the corresponding tensile strength at each case of He/V ratio.

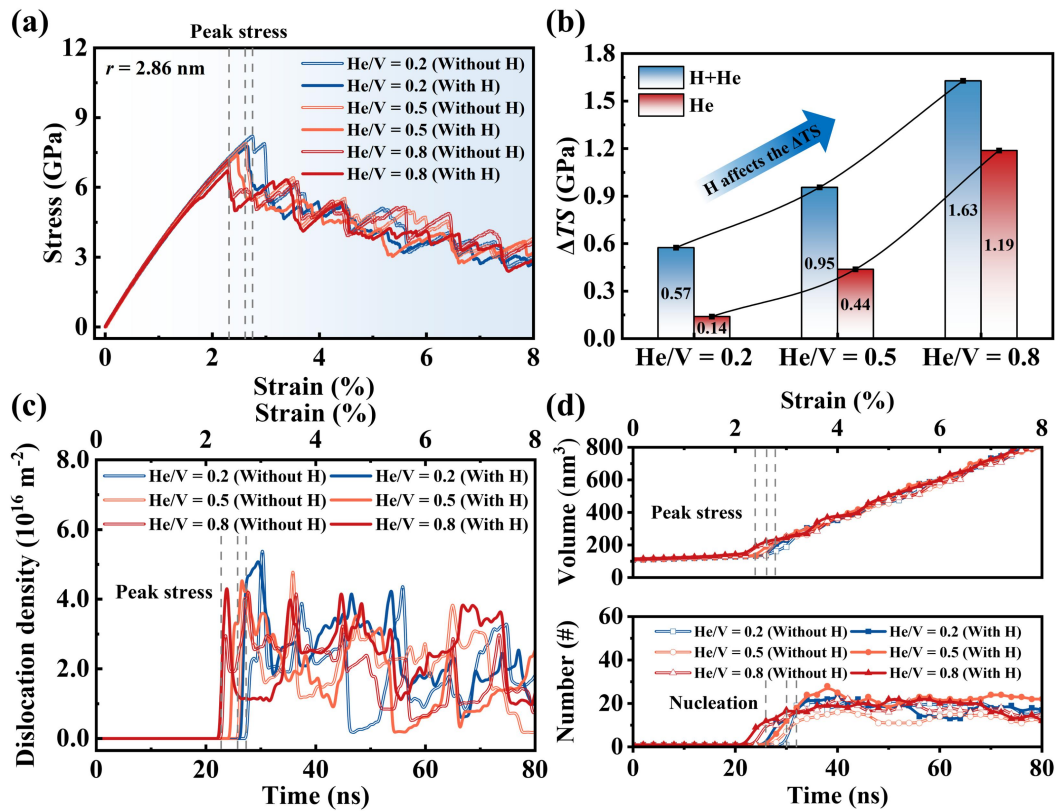

**Fig. 4.** (a) Stress-strain curve response for cavities containing both H and He atoms compared to those containing only He atoms at $r$ = 2.86 nm. (b) The values of tensile strength variation (Δ$TS$) for cavities with different structures are compared to those of corresponding-sized pure voids. Evolution of the dislocation density (c) and (d) the volume and nucleation number for the above cavities (with or without H).

In addition, before reaching peak stress, the strain field around the cavity is also enhanced by the H within the cavity during the elastic deformation stage, as demonstrated in Figs. 5 (a1-a4). Unlike the behavior of He atoms (trapped within the core), the strain field induced by cavity defects is primarily influenced by H atoms trapped at the surface by the cavity. As demonstrated in Figs. 4 (c) and 5 (c1-c4), during the plastic damage phase, both cavities containing H and He atoms and those containing only He atoms exhibit emission dislocations that promote the growth of the initially formed cavities. Based on the presence of He, the dominant role of H atoms in causing plastic damage to the cavity seems not to be altered, but rather further lowers the strain threshold for the cavity to emit dislocations. The mechanism by which H is present on the surface of a cavity (or rather the He$_n$V$_m$ complex) promotes dislocation emission from the cavity under stress/strain fields is similar to that of H acting at other defects, such as crack tips and grain boundaries [56–59]. This behavior is consistent with the Hydrogen Enhanced Localized Plasticity (HELP) mechanism, wherein hydrogen-defect interactions facilitate deformation. [60–62]. A detailed analysis of this mechanism will be provided in a subsequent study.

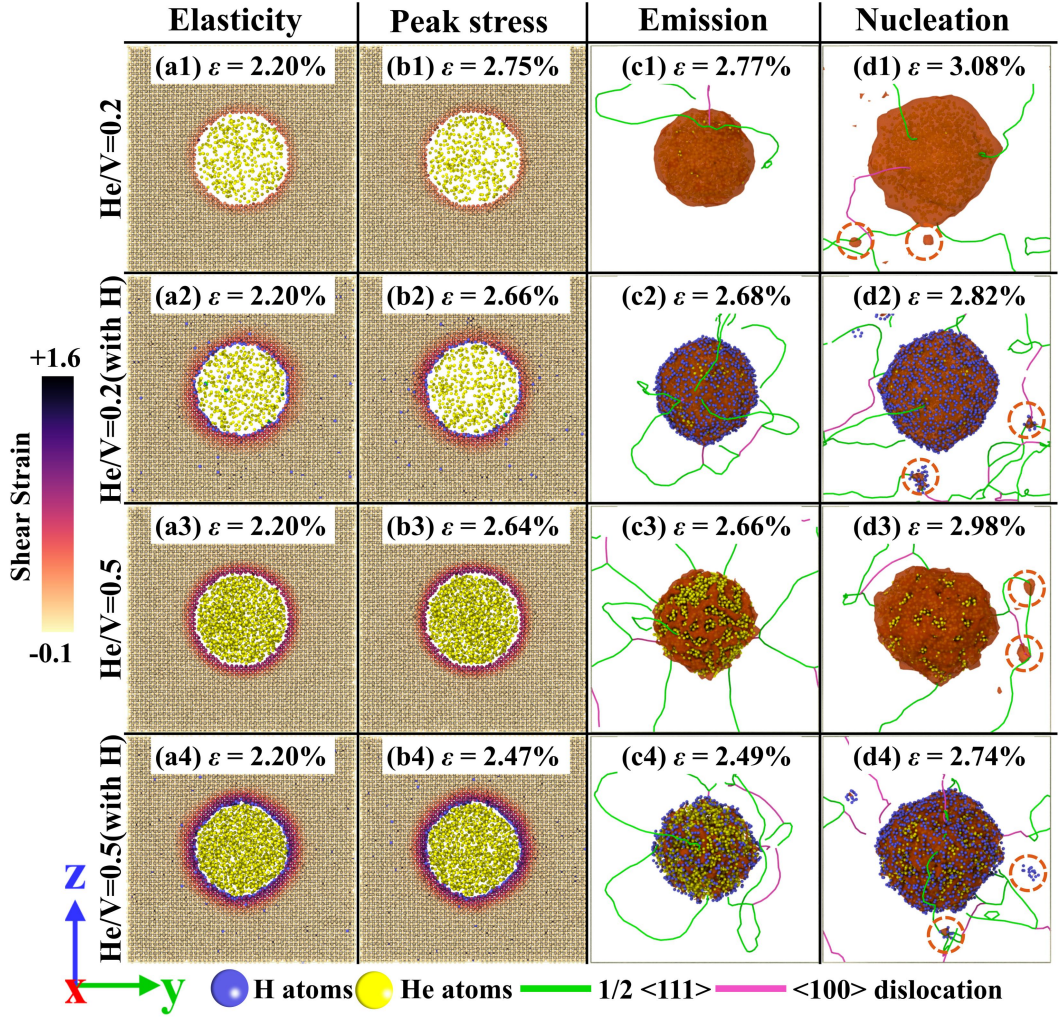

**Fig. 5.** Atomic distribution (emphasized by shear strain) and defect configuration diagrams (cavity surfaces characterized as orange by the CSM algorithm) for cavities with a radius of 2.86 nm and He/V ratios of 0.2 and 0.5 (with or without H) under a hydrostatic strain ($\varepsilon$) field. The cross-sectional views of the two He/V ratios correspond to the deformation response stages indicated by points (a), (b), (c), and (d) on the stress-strain curve shown in Fig. 4 (a). The cavities nucleated in the BCC-Fe during the plastic deformation stage are marked with orange circles.

With continued dislocation emission from cavity surfaces, the number of newly nucleated cavities in the matrix initially increases before eventually stabilizing, as shown in Figs. 4 (d) and 5 (d1-d4). Under applied stress, the presence of H consistently promoted cavity nucleation and reduced the corresponding strain threshold in the matrix material across all He/V ratios investigated. Due to the high binding energy of He atoms to the cavity (~2.0 eV) [14,63], He atoms remain stably trapped within the cavity core even at current conditions. Conversely, due to the lower binding energy between H and the cavity [14,64,65], a portion of the H atoms trapped on the surface desorb into the bulk. This process facilitates that these mobile H atoms promote the formation of additional vacancies under the local stress field [45]. Subsequently, the total number of bulk cavities is simultaneously regulated by two mechanisms: H-enhanced strain-induced vacancy [45,58] and H-influenced cavity growth and coalescence [66,67]. Therefore, during the plastic damage stage, the number of nucleated voids (cavities) within the sample is also affected by H atoms desorbed from the cavity surfaces.

In conclusion, the dynamic response under triaxial/hydrostatic tensile loading is significantly altered by the cavities with core-shell structures. During the elastic deformation, H and He atoms considerably enhance the stress field around the cavities. At the elastoplastic transition, the tensile strength of the bulk material is effectively reduced by H and He. In the subsequent plastic stage, these cavities primarily promote cavity growth through dislocation emission and concurrently facilitate the nucleation of new cavities within the matrix. Within this structure, H and He exhibit synergistic damage effects, with H playing a role mechanistically analogous to that of He. The deformation of cavities with a core-shell structure is driven primarily by He-generated internal pressure, while H segregated at the cavity surface modifies the interfacial energy and structure, thereby accelerating deformation. The role of H in cavities (containing both H and He atoms) across various sizes and He/V ratios is observed to align with a HELP-dominated mechanism. Furthermore, further studies will focus on investigating the mechanical response and defect evolution of cavities containing H and He atoms under varying conditions, such as temperature, strain rate, and H/He implantation rate.

**Declaration of competing interest**

The authors declare that they have no known competing financial interests or personal relationships that could have appeared to influence the work reported in this paper.

## Acknowledgments

This work was supported by the National MCF Energy R&D Program (Grant No. 2022YFE03110000); the National Natural Science Foundation of China (Grant Nos. 12192280, 12422514, 12275009, and 11935004); and the National Key R&D Program of China under Grant No. 2025YFB3003603. The authors gratefully acknowledge the computing resources and support provided by the High-Performance Computing Platform of the Center for Life Science (Peking University).